\documentstyle[preprint,aps,eqsecnum]{revtex}

\newcommand\ignore[1]{}

\begin{document}

\draft
\title{MOLECULAR CONDUCTORS WITH TWO-CHAIN ORBITALS}
\author{H. Tajima, J. Shiraishi, and M. Kohmoto}
\address{Institute for Solid State Physics,
University of Tokyo, Roppongi, Minato-ku, Tokyo 106, Japan}

\maketitle
\begin{abstract}
We investigate a theoretical model applicable to molecular conductors, such
as TTP and $ \rm M(dmit)_2$ salts [M=Ni, Pd, Pt], whose charge carriers orig
inate
from two kinds of molecular orbitals. The model Hamiltonian consists of two
independent Hubbard chains. The degree of charge transfer between the two
chains is calculated. The results qualitatively agree with some experimental
data.

\end{abstract}

\narrowtext
\newpage

\section{Introduction}
In most of the molecular conductors which exhibit metallic properties at low
temperatures, charge carriers originate from molecular orbitals of one kind
such as the highest occupied molecular orbital (HOMO) of a donor molecule or
the lowest unoccupied molecular orbital (LUMO) of an acceptor molecule.[1]
For example, the charge carriers of $\rm \beta$-$\rm (BEDT$-$\rm TTF)_2 I_3$
originate from the
HOMO's of BEDT-TTF molecules while $\rm I_3^-$ anions merely work
as reservoirs of
negative charges. In this type of molecular conductors, the chemical
composition in principle determines the filling factor of the conduction band.

On the other hand, charge carriers in some molecular conductors originate
from two kinds of molecular orbitals (``two-orbital'' molecular conductors).
Here we call one of the two kinds of molecular orbitals, having the larger
orbital energy, as the upper orbital (UO), and the other, having the lower
orbital energy, as the lower orbital (LO). The energy separation between the
UO and LO is denoted by $\Delta$.

A molecular conductor composed of segregated columns of donor molecules and
columns of acceptor molecules, such as TTF$\bullet$TCNQ, may be considered as a
two-component type of the "two-orbital" molecular conductor. The UO and LO
in this type are the LUMO of an acceptor molecule and the HOMO of a donor
molecule, respectively.

Another type of the ``two-orbital'' molecular conductor is a single-component
one, where the UO and LO come from molecules of one kind. The salt of
$\rm TTF[Ni(dmit)_2]_2$ is classified into this type. The UO and LO in this
salt are
the HOMO and LUMO of $\rm Ni(dmit)_2$, respectively.[2] For this type, one can
estimate $\Delta$ approximately from the intra-molecular optical absorption
for the
molecule which possesses the UO and LO. The values of $\Delta$, thus determi
ned for
typical organic donors and acceptors, are listed in Table 1.
In the ``two-orbital'' molecular conductors, the charge transfer between the
two molecular orbitals is a controllable parameter, which one can expect to
enrich their properties. In order to reveal them, we have studied a
two-chain model.

\section{Model Hamiltonian}

The Hamiltonian which describes two-chained molecular orbitals interacting
through the on-site Coulomb repulsion is written as
\begin{eqnarray}
H&=&H_u+H_l+T+Q,   \nonumber \\
H_u&\equiv&
\Delta \sum_{i,\sigma}a_{i\sigma}^\dagger a_{i\sigma}
-t_{uu}\sum_{i,\sigma}
\left(a_{i\sigma}^\dagger a_{i+1\sigma}+{\rm c.c.}\right)
+U_{uu}\sum_{i }
 a_{i\uparrow}^\dagger a_{i\uparrow}
 a_{i\downarrow}^\dagger a_{i\downarrow} , \label{twohub}\\
H_l&\equiv&
-t_{ll}\sum_{i,\sigma}
\left(b_{i\sigma}^\dagger b_{i+1\sigma}+{\rm c.c.}\right)
+U_{ll}\sum_{i }
 b_{i\uparrow}^\dagger b_{i\uparrow}
 b_{i\downarrow}^\dagger b_{i\downarrow} ,    \nonumber\\
T&\equiv&
-t_{ul}\sum_{i,\sigma}
\left(a_{i\sigma}^\dagger b_{i+1\sigma}+{\rm c.c.}\right)
-t_{lu}\sum_{i,\sigma}
\left(b_{i\sigma}^\dagger a_{i+1\sigma}+{\rm c.c.}\right),    \nonumber\\
Q&\equiv&
U_{ul}\sum_{i,\sigma,\sigma' }
 a_{i\sigma}^\dagger a_{i\sigma}
 b_{i\sigma'}^\dagger b_{i\sigma'} , \nonumber
\end{eqnarray}
where $a_{i\sigma}^\dagger$ and $a_{i\sigma}$
are the creation and annihilation operator of an electron at the
$i$-th site with spin $\sigma$  in the UO;
$b_{i\sigma}^\dagger$ and $b_{i\sigma}$ are those in the LO;
$t_{uu}$, $t_{ll}$, $t_{ul}$,
$t_{lu}$ are transfer integrals; and $U_{uu}$, $U_{ll}$,
$U_{ul}$ are the on-site Coulomb
repulsions. The model Hamiltonian (1) consists of two Hubbard chains
($H_{uu}$, $H_{ll}$) and their couplings ($T$, $Q$).
If the couplings are neglected, the problem
is reduced to the two Hubbard models which can be solved by Bethe ansatz.
this approximation may be partially justified by the following considerations:

First, $T$ is anticipated to be unimportant until the charge transfer from the
LO to the UO occurs, since $t_{ul}$, $t_{lu}$
are smaller than $t_{uu}$, $t_{ll}$, and
negligible compared to $\Delta$.

Secondly,
\begin{eqnarray}
Q&=&
{U_{ul}\over 2}\sum_{i  }
\left(
 a_{i\uparrow}^\dagger a_{i\uparrow}+
 a_{i\downarrow}^\dagger a_{i\downarrow}+
 b_{i\uparrow}^\dagger b_{i\uparrow}+
 b_{i\downarrow}^\dagger b_{i\downarrow}\right)^2  \\
&&-{U_{ul}\over 2}\sum_{i ,\sigma }
\left( a_{i\sigma}^\dagger a_{i\sigma}+
 b_{i\sigma }^\dagger b_{i\sigma } \right)
-U_{ul}\sum_{i  }
\left(
 a_{i\uparrow}^\dagger a_{i\uparrow}
 a_{i\downarrow}^\dagger a_{i\downarrow}+
 b_{i\uparrow}^\dagger b_{i\uparrow}
 b_{i\downarrow}^\dagger b_{i\downarrow}\right),  \nonumber
\end{eqnarray}
where the first term is proportional to the fluctuation of electron density
at each site; the second term is proportional to the total electron number;
and the third term is on-site Coulomb repulsion, which simply reduces the
effective value of $U_{uu}$ in $H_u$ and that of $U_{ll}$ in $H_l$.
Consequently, $Q$ can be
neglected until the fluctuation of electron density becomes dominant.

Since $H_u$ and $H_l$ are now independent, the ground state is obtained by
imposing the same chemical potential to the two chains. More explicitly one
can obtain the ground state energy by minimizing
$E\equiv E[{\rm  (UO)}^{N_u}]+E[{\rm  (LO)}^{N_l}]$ under the condition of
$N_u+N_l=N_t$, where $E[{\rm  (UO)}^{N_u}]$
denotes the ground state energy for $H_u$ when $N_u$ electrons
are in the UO, and $E[{\rm  (LO)}^{N_l}]$
denotes the corresponding energy for $H_l$ when $N_l$
electrons are in the LO. The total electron number in the system is $N_t$.

In the following calculation, we use the ground state energy per one site
for the Hubbard chain. This energy,
$\varepsilon_{Hub.}(u,n)\equiv E/N_s$, was originally derived by Lieb and
Wu:[11,12,13]
\begin{eqnarray}
\varepsilon_{Hub.}(u,n)&=&-2t \int_{-Q}^{Q}dk\;\cos k \;\rho(k),\nonumber\\
\int_{-Q}^{Q}dk\; \rho(k)&=&N/N_s, \label{baeq} \\
2 \pi \rho(k)&=&1+\cos k \int_{-Q}^{Q}dk'\; \rho(k')
{8 \pi \over u}R\left( {4(\sin k-\sin k') \over u}\right), \nonumber \\
R(x)&=&{1 \over 4 \pi}
\int_{-\infty}^{\infty} {{\rm sech}(\pi t/2)\over 1+(x+t)^2}dt,\nonumber
\end{eqnarray}
where $N_s$ , $N$ ($<N_s$) are the total number of sites and electrons;
$n\equiv N/N_s$; and
$u = U/t$, respectively. By solving these equations we can study the range
$n<1$. For the range $n>1$, $\varepsilon_{Hub.}(u,n)$
 is given by an identical equation,
\begin{equation}
\varepsilon_{Hub.}(u,n)=\varepsilon_{Hub.}(u,2-n)+U(n-1),
\end{equation}
coupled with Eq.(\ref{baeq}).
Since a closed form of (\ref{baeq}) is not known, we have obtained following
functions, $\varepsilon_1(u,n)$ and $\varepsilon_2(u,n)$
which approximate $\varepsilon_{Hub.}(u,n)$:
\begin{eqnarray*}
  \varepsilon_{Hub.}(u,n) &\cong& \varepsilon_1(u,n)\quad    (0<n<1)\\
               &\cong& \varepsilon_2(u,n)\quad         (1<n<2)
\end{eqnarray*}
within the ansatz
\begin{eqnarray*}
\varepsilon_1(u,n)&=& -A(u)\;t\;
\sin(B(u)n+C(u)n^2+D(u)n^3);\\
\varepsilon_1(u,n)&=&\varepsilon_1(u,2-n)+U(n-1);
\end{eqnarray*}
\begin{eqnarray}
A(u)&=&{2 \over \pi}\left( 1+(1+P_4 u+P_5 u^2)
\exp(-P_6 u)\right) \quad (0<u<10), \nonumber\\
A(u)&=&{2 \over \pi}+P_{13}u^{-P_{14}} \quad (10<u<250); \nonumber\\
B(u)&=&2/A(u); \\
C(u)&=&P_7 u/(1+P_8 u^2+P_9 u^3)+P_{10} u \quad (0<u<10), \nonumber \\
C(u)&=&P_{15} u^{-P_{16}}+P_{17} u^{-P_{18}} \quad (10< u< 250); \nonumber\\
D(u)&=& (\pi/2-B(u)n_{\rm max}-C(u)n_{\rm max}^2)/n_{\rm max}^3; \nonumber
\end{eqnarray}
where
\begin{eqnarray*}
n_{\rm max}(u)&=&{1 \over 2} \left(
 1+(1+P_1 u+P_2 u^2)
\exp(-P_3 u)\right) \quad(0<u<10),\\
n_{\rm max}(u)&=&{1 \over 2} (1+P_{11}u^{-P_{12}})\quad  (10< u < 250);
\end{eqnarray*}
$$
\begin{array}{lll}
P_{1} = -0.08705;& P_{2} = 0.01366; &P_{3} = 0.23603;\\
P_{4} = -0.10512;& P_{5} = 0.0207;& P_{6} = 0.28767;\\
P_{7} = -0.14159; &P_{8} = 0.06386;& P_{9} = 0.00582;\\
P_{10} = -0.00903;& P_{11}= 0.76063;& P_{12}= 1.03065;\\
P_{13}= 0.80627;& P_{14}= 1.03467;& P_{15}= -1.3295;\\
P_{16}= 0.75014;& P_{17}= 0.12461;& P_{18}= 0.39471.
\end{array}
$$
The deviations of $\varepsilon_1(u,n)$ and
$\varepsilon_2(u,n)$ from $\varepsilon_{Hub.}(u,n)$ are
about $0.001t$ in the range of $0< u<250$.

\section{Results}

We take $U_{uu}=U_{ll}=U$ and $t_{uu}=t_{ll}=t$ for simplicity.
Generalization to the other
cases is straightforward.\\

\noindent
a) ${\rm (UO)}^{(1.5+x)N_s}{\rm (LO)}^{(2-x)N_s}$

This is the case corresponding to the charge-transfer salts of
$\rm TTP^{0.5+}$.[9]
The parameter $x$ represents the degree of charge transfer from the LO to the
UO. The ground-state energy per one site,
$\varepsilon(\equiv  E/N_s)$, is expressed by
\begin{eqnarray}
\varepsilon&=&
\Bigl(\varepsilon_2(U/t,1.5+x)+(1.5+x)\Delta \Bigr) +
\varepsilon_2(U/t,2-x),\\
&=&1.5 U + (1.5+x)\Delta +\varepsilon_1(U/t,0.5-x)+
\varepsilon_1(U/t, x). \nonumber
\end{eqnarray}
In what follows, $x$ is determined as the value which minimizes e for a given
set of $U/t$ and $\Delta/t$:

\noindent
i)      If  ${\partial \varepsilon \over \partial x} \Bigl|_{x=+0}>0$,
        we have, $x=0$.

\noindent
ii)     Next, if
${\partial \varepsilon \over \partial x} \Bigl|_{x=+0}<0$,
        we obtain $x$ as the solution of
\begin{eqnarray}
{\partial \varepsilon \over \partial x}
= \Delta-
{\partial \varepsilon_1(U/t,n) \over \partial n} \Biggl|_{n=0.5-x}+
{\partial \varepsilon_1(U/t,n) \over \partial n} \Biggl|_{n= x}
=0.
\end{eqnarray}

Figure 1 shows the degree of charge transfer thus obtained. In the case of
TTP salts,[9] the value of $\Delta/t$ is considered to be always larger than
2.[10,14] Therefore we conclude that the charge transfer from the LO to the
UO does not occur in these salts. In fact, properties of these salts are
consistent with the expectation given by the extended H\"ukel calculation
where only the HOMO is taken into account.[14]\\

\noindent
b) ${\rm (UO)}^{(1+x)Ns}{\rm (LO)}^{(2-x)N_s}$

This is the case corresponding to the charge-transfer salts of
$\rm TTP^+$[9] and
$\rm [M(dmit)_2]^-$ (M=Ni, Pd, Pt).
The ground-state energy per site is expressed by
\begin{eqnarray}
\varepsilon&=&U+(1+x)\Delta+ \varepsilon_1(U/t,1-x)+
\varepsilon_1(U/t, x) .
\end{eqnarray}
The degree of charge transfer, $x$, is obtained as follows:

\noindent
i)      If ${\partial \varepsilon \over \partial x} \Bigl|_{x=+0}>0$,
        we have $x=0$.

\noindent
ii)     Next if ${\partial \varepsilon \over \partial x} \Bigl|_{x=+0}<0$,
        we obtain $x$ as the solution of
\begin{equation}
 \Delta-
{\partial \varepsilon_1(U/t,n) \over \partial n} \Biggl|_{n=1-x}+
{\partial \varepsilon_1(U/t,n) \over \partial n} \Biggl|_{n= x}
=0.
\end{equation}

Figure 2 shows the degree of charge transfer thus obtained. The ground state
of the one-dimensional Hubbard model is insulating at half filling.
Therefore, the two-chained Hubbard model (\ref{twohub})
predicts an insulating state at
$x=0$, metallic state for $x\neq 0$, and metal-insulator transition on the b
oundary
of these two regions.

In the case of $\rm TTP^+$[9] salts, the value of $U/t$ is about 3.[10] Thus
, from
Fig 2a, one may expect that metallic state in these salts if $\Delta/t<3$. This
condition is not unreasonable. In fact, Mori et al recently found the
metallic behavior above 160 K in TTM-TTP$\bullet {\rm I}_3$, which is a
charge-transfer salt
of TTM-$\rm TTP^+$. Although they did not consider the possibility of the charge
transfer between the LO and UO[15], our calculation suggests that such
charge transfer causes the metallic behavior of this salt above 160 K.\\

\noindent
c) ${\rm (UO)}^{(0.5+x)N_s}{\rm (LO)}^{(2-x)N_s}$

This is the case corresponding to the charge-transfer salts of
$ \rm [M(dmit)_2]^{0.5-}$ (M=Ni, Pd, Pt).
The ground-state energy per one site, $\varepsilon$, is
expressed as follows:
\begin{eqnarray}
\varepsilon&=&U(1-x)+(0.5+x)\Delta+
 \varepsilon_1(U/t,0.5+x)+
\varepsilon_1(U/t, x)\qquad (0<x<0.5),\\
\varepsilon&=&0.5 U +(0.5+x)\Delta+
 \varepsilon_1(U/t,1.5- x)+
\varepsilon_1(U/t, x)\qquad (0.5<x<0.75).
\end{eqnarray}
The degree of charge transfer, $x$, is determined as follows:

\noindent
i)      If ${\partial \varepsilon \over \partial x} \Bigl|_{x=+0}>0$,
        we have $x=0$.

\noindent
ii)     If
${\partial \varepsilon \over \partial x} \Bigl|_{x=+0}<0
 {\partial \varepsilon \over \partial x} \Bigl|_{x=0.5-0}$,
        we obtain $x$ as the solution of
\begin{equation}
-U+\Delta+
{\partial \varepsilon_1(U/t,n) \over \partial n} \Biggl|_{n=0.5+x}+
{\partial \varepsilon_1(U/t,n) \over \partial n} \Biggl|_{n= x}
=0.
\end{equation}

\noindent
iii)    If
${\partial \varepsilon \over \partial x} \Bigl|_{x=0.5-0}<0<
 {\partial \varepsilon \over \partial x} \Bigl|_{x=0.5+0}$,
        we have $x=0.5$.

iv) If ${\partial \varepsilon \over \partial x} \Bigl|_{x=0.5+0}<0$,
        we obtain $x$ as the solution of
\begin{equation}
 \Delta-
{\partial \varepsilon_1(U/t,n) \over \partial n} \Biggl|_{n=1.5-x}+
{\partial \varepsilon_1(U/t,n) \over \partial n} \Biggl|_{n= x}
=0.
\end{equation}

Figure 3 shows the degree of charge transfer thus obtained. It should be
noted that the state $x=0.5$ is stable in this model, as can be seen from Fig.
3a. This comes from the fact that
$\partial \varepsilon_{Hub.}(u,n)/\partial n$ is discontinuous at $n=1$.

The present model Hamiltonian does not include any contribution from the
electron-phonon interaction. The interaction is expected to make the state,
$x=0.5$, more stable by forming a gap in the middle of the half-filled UO band
(Peierls transition). In this sense, the dimerized structure frequently
found in $\rm Pd(dmit)_2$ and $\rm Pt(dmit)_2$
salts [16] may be considered as a result of
Peierls transition.\\

\noindent
d) ${\rm (UO)}^{xN_s}{\rm (LO)}^{(2-x)N_s}$

This is the case corresponding to the "two-orbital" molecular conductors of
the two-component type, such as $\rm TTF\bullet TCNQ$.
The ground-state energy per one
site, $\varepsilon$, is expressed by
\begin{eqnarray}
\varepsilon&=&U(1-x)+x\Delta+
 2\varepsilon_1(U/t, x).
\end{eqnarray}
The degree of charge transfer, $x$, is determined as follows:

i)      If ${\partial \varepsilon \over \partial x} \Bigl|_{x=+0}>0$,
        we have $x=0$.

ii)     If
${\partial \varepsilon \over \partial x} \Bigl|_{x=+0}<0<
 {\partial \varepsilon \over \partial x} \Bigl|_{x=1-0}$,
        we obtain $x$ as the solution of
\begin{equation}
-U+\Delta+
{\partial \varepsilon_1(U/t,n) \over \partial n} \Biggl|_{n=0.5+x}+
{\partial \varepsilon_1(U/t,n) \over \partial n} \Biggl|_{n= x}
=0.
\end{equation}

iii)    If ${\partial \varepsilon \over \partial x} \Bigl|_{x=1- 0}<0$,
        we have $x=1$.

Figure 4 shows the degree of charge transfer thus obtained. This figure
exhibits an insulating state at $x=0$ and $x=1$; metallic state for $0<x<1$.
Although a ``two-orbital'' molecular conductor of a single-component type
which exhibits metallic behavior has not been found, the present model
predicts that synthesis of such compound is possible by choosing appropriate
values of $U/t$ and $\Delta/t$.

\section{Conclusion}

We found that the on-site coulomb interaction plays quite an important role
in the ``two-orbital'' molecular conductors. The effect of this interaction
grows the electron density in the UO becomes smaller. The model gives a
plausible explanation for the metallic behavior of
$\rm TTM$-$\rm TTP\bullet I_3$ and the
strongly dimerized structure frequently observed in $\rm Pd(dmit)_2$
and $\rm Pt(dmit)_2$
salts. In spite of its simplicity, this model is versatile and applicable to
many molecular conductors.\\

\noindent
{\sl Acknowledgment}

\noindent
We deeply thank Prof. Kyuya Yakushi , Prof.
Reizo Kato and Dr. David Lidsky  for
valuable discussions. This work was supported by the Grant-in-Aid for the
Special Project Research on the "Novel Electronic State in Molecular
Conductors" and by the Grant-in-Aid for Scientific research (No. 08640733)
from the Ministry of Education, Science and Culture, Japan.

\begin{table}
s.:determined from an absorption spectrum of solution\\
c.:determined from an absorption spectrum of a single crystal\\

\begin{tabular}{ l l l l l }
                   &$\Delta$(eV)  &UO    &LO       &  Ref.   \\
$acceptor$         &              &      &         &         \\
$\rm TCNQ^-$       &1.3           &LUMO  &HOMO     &[3]${}^{s.}$\\
$\rm Ni(dmit)_2^-$ &1.1           &LUMO  &HOMO     &[4]${}^{s.}$\\
$\rm Pd(dmit)_2^-$ &0.9           &LUMO  &HOMO     &[4]${}^{s.}$\\
                   &              &      &         &  \\
$donor$            &              &      &         &  \\
$\rm TTF^+$        &2             &HOMO  &2nd HOMO &[5]${}^{s.}$\\
BPDT-$\rm TTF^+$   &1.7           &HOMO  &2nd HOMO &[6]${}^{s.}$\\
BMDT-$\rm TTF^+$   &1.4           &HOMO  &2nd HOMO &[7]${}^{c.}$\\
BEDT-$\rm TTF^+$   &1.4           &HOMO  &2nd HOMO &[8]${}^{c.}$\\
$\rm TTP^+$  [9]   &0.9           &HOMO  &2nd HOMO &[10]${}^{c.}$\\
\end{tabular}
\caption{
The energy separation between the UO and LO. The molecular orbitals
(LUMO, HOMO, and  2nd HOMO) are defined in the neutral state of an isolated
molecule.}
\end{table}

\begin{figure}
\figure
Fig. 1 (a)The degree of charge transfer ($x$) obtained for the two-chained
orbitals, ${\rm (UO)}^{(1.5+x)N_s}{\rm (LO)}^{(2-x)N_s}$.
(b)The regions for $x=0$ and for $0<x<0.25$.

\figure
Fig. 2 (a)The degree of charge transfer ($x$) obtained for the two-chained
orbitals, ${\rm (UO)}^{(1+x)N_s}{\rm (LO)}^{(2-x)N_s}$.
(b)The regions for $x=0$ and for $0<x<0.5$.

\figure
Fig. 3 (a)The degree of charge transfer ($x$) obtained for the two-chained
orbitals, ${\rm (UO)}^{(0.5+x)N_s}{\rm (LO)}^{(2-x)N_s}$.
(b)The regions for $x=0$; $0<x<0.5$;
$x=0.5$;
and $0.5<x<0.75$.

\figure
Fig. 4 (a)The degree of charge transfer ($x$) obtained for the two-chained
orbitals, ${\rm (UO) }^{xN_s}{\rm (LO)}^{(2-x)N_s}$.
(b)The regions for $x=0$; $0<x<1$; and $x=1$.

\end{figure}


\begin{references}
\bibitem{TTP}
In this paper we use ``TTP'' as a generic name of BDT-TTP and its
derivatives, such as TTM-TTP, TMET-TTP.
\end{references}
\end{document}